# Optimization of Moment Masking for CO Spectral Line Surveys

T. M. Dame
Harvard-Smithsonian Center for Astrophysics
60 Garden Street, Cambridge, MA 02138
tdame@cfa.harvard.edu

ABSTRACT

We describe and refine the masked moment method used to suppress noise in Galactic and extragalactic spectral line surveys. By applying the refined technique to an essentially noise-free CO molecular cloud survey with Gaussian noise added, we determine the optimum masking parameters for typical CO surveys such as those presented in Dame et al. (2001).

1. Introduction

The CO surveys presented in Dame et al. (2001) are 3-dimensional (v-x-y) and typically contain hundreds of spectral channels at each of thousands of independent (x,y) sky positions. The simplest reduction of such a data cube is to sum over one of the axes to produce a zeroth moment map, either a velocity-integrated sky map or a position-velocity map. While straightforward and unbiased, this approach will substantially degrade the signal-to-noise ratio of typical emission features, which extend over just a small fraction of the spectral channels and sky positions. In some cases small clouds that have high significance in single-channel maps will fall below the noise in the integrated map.

A common remedy for this problem is "clipping", in which all spectral channels with intensities below some statistical significance level, typically 3 times the rms noise, are set to zero before summing. This technique is very simple to implement and can be remarkably effective in defining clouds for qualitative analysis. However, at every reasonable clipping level the technique suffers from dueling biases, a positive one owing to noise spikes extending above the clipping level and a negative one owing to the clipping of both spectral line wings and weak, extended emission. One can eliminate the positive bias by clipping absolute intensities, but this only forces the clipping level higher to avoid unacceptable noise, which in turn worsens the negative bias. We will show below that for a CO survey of a typical molecular cloud such as that in Taurus, there is no clipping level at which most of the noise is suppressed and most of the emission is maintained.

A shortcoming of clipping is that it determines whether or not a particular spectral channel contains real emission based only on the intensity level in that channel, whereas a seasoned radio astronomer would, in addition, look for spatial and velocity coherence of the signal. Essentially, moment masking is a refinement of the clipping method in which the coherence of the signal—determined from a smoothed version of the data cube—is considered in determining which peaks are real (and consequently not blanked). It is important to emphasize that the smoothed cube is used only to identify and blank certain regions of the raw cube that are likely emission-free. The resulting moment-masked cube maintains its full angular and velocity resolutions, and for sufficiently high masking levels it is unbiased.

With appropriate choices of the smoothing and masking parameters, moment masking will reliably extract all of the emission lines that one would judge by eye to be real, as well as many less conspicuous lines that are statistically significant in groups owing to their spatial and velocity coherence. Further, we will show below that for a typical cloud survey there is a comfortable range of masking parameters in which most of the noise is removed from the integrated maps and nearly all of the emission is maintained.

The main drawback of moment masking is that the noise in the integrated maps is not constant, since different numbers of channels are summed over at each point. Also, since the masking generally produces integrated maps with large areas set to zero, any noise or data artifacts that exceed the masking level will produce map features that appear to the eye as highly significant. Given these limitations, moment masking requires data cubes that have fairly uniform noise as well as uniform sampling.

## 2. History

The moment masking technique was developed to analyze early extragalactic 21 cm observations (e.g., Rots et al. 1990). Moment masking can be performed with the routine MOMNT in the widely-used Astronomical Image Processing System (AIPS):

http://www.aips.nrao.edu/cgi-bin/ZXHLP2.PL?MOMNT

although it cannot be used for the refined version of the technique proposed here. Other papers that discuss and/or use the technique include Tilanus & Allen (1991) for HI and M51, Adler et al. (1992) for CO in M51, Digel et al. (1996) for CO toward W3, Loinard et al. (1999) for CO in M31, and Dame et al. (2001). In all of these papers, however, moment masking is merely mentioned in passing, or at most discussed in a general way in one or two paragraphs. One of the motivations for the present paper is to give a fuller description of the technique particularly as it was applied to the CfA CO surveys, so that others may cite, use, and possibly improve the technique.

## 3. Procedure

We consider a spectral line data cube $T(v,x,y)$ for which we want to produce a moment-masked version $T_M(v,x,y)$ in which most of the emission-free pixels are blanked. Here v is velocity and (x,y) are the sky coordinates. With few exceptions the surveys in the CfA CO surveys have a velocity spacing dv of either 0.65 or 1.3 km s$^{-1}$ and a spatial spacing ds of either 1/8° (~1 beamwidth) or 1/4°.

i. Determine the rms noise in T(v,x,y).

ii. Generate a smoothed version of the data cube $T_S$(v,x,y) by degrading the resolution spatially by a factor of 2 and in velocity to the width of the narrowest spectral lines generally observed. For the CfA CO surveys we used convolution with a Gaussian smoothing kernel with $fwhm_s$ = 1/4° (or 1/2°) and $fwhm_v$ = 2.5 km s$^{-1}$.

iii. Determine the rms noise in $T_S$(v,x,y). It is essential that the rms be measure accurately, either by using regions expected to be emission-free (e.g., beyond the terminal velocity in the inner Galaxy, or at high Galactic latitude and high LSR velocity) or by iteratively excluding regions of significant emission. Take care that your smoothing algorithm does not zero (rather than blank) edge pixels since this would artificially lower the rms. Likewise be aware of under-sampled regions that were filled by linear interpolation, since these will have higher rms in the smoothed cube.

iv. Let the clipping level $T_c$ equal 5 times the rms noise in the smoothed cube; the factor 5 will be discussed and justified below.

v. Generate a masking cube M(v,x,y) initially filled with zeros with the same dimensions as T and $T_S$. The moment masked cube $T_M$(v,x,y) will be calculated as M*T.

vi. For each pixel $T_S(v_i, x_j, y_k)$ > $T_c$, unmask (set to 1) the corresponding pixel in M.

vii. Also unmask all pixels in M within the smoothing kernel of $T_S(v_i, x_j, y_k)$, since all of these pixels weigh into the value of $T_S(v_i, x_j, y_k)$. That is, unmask within nv pixels in velocity and within ns pixels spatially, where

$$nv = 0.5*fwhm_v / dv$$
$$ns = 0.5*fwhm_s / ds$$

For a typical CO cube with dv = 1.3 km s$^{-1}$ and ds = 1/8°, nv = 1 and ns = 1. For factor-2 smoothing in each dimension all pixels adjacent to $T_S(v_i, x_j, y_k)$ are unmasked, a total of 3 x 3 x 3 pixels. This step is a refinement of conventional moment masking.

viii. Calculate the moment-masked cube $T_M$(v,x,y) = M*T.

ix. Zeroth moment maps can be obtained by summing over any dimension of $T_M$(v,x,y).

4. Refinements of the Method

It is worth noting that step vii above is a refinement of the basic method in which only one channel in T is unblanked for every channel above the clipping level in $T_S$ (step vi). The necessity of step vii stems from the fact that a peak at $T_S(v_i, x_j, y_k)$ implies only that there is significant emission *somewhere* within the smoothing kernel of $T(v_i, x_j, y_k)$. That emission need not peak at $T(v_i, x_j, y_k)$, nor does there even need to be *any* emission at $T(v_i, x_j, y_k)$. Consider, for example, the case of a small annular shaped cloud with a radius of about ds centered on $T(v_i, x_j, y_k)$. Such a geometry could lead to a situation in which only the pixel $T_S(v_i, x_j, y_k)$ is above the clipping level, but there is no emission at all at $T(v_i, x_j, y_k)$. Conventional moment masking would completely miss this cloud, while the modification in step vii would unmask most or all of it. This refinement is fairly innocuous in that it simply unmasks a extra layer (or two) of pixels around any volume of the cube with significant emission, allowing a small amount of additional noise into $T_M$.

Choosing $T_c$ at 5-sigma differs from the conventional method in which 3-sigma is generally adopted. A higher significance level assures that every pixel above the clip level in $T_S$ corresponds to significant emission in T, and step vii assures that most of that real emission will be unmasked in $T_M$. Owing to the several times greater sensitivity of the smoothed cube to clouds at least as large as the smoothing kernel, clipping at 5 sigma in $T_S$ will unmask all such clouds in T that have mean line intensities above ~2 sigma.

5. Optimizing the Method

Here we will start with an essentially noiseless map of a typical molecular cloud, then add a substantial amount of Gaussian noise to the data cube, then test how well we can recover the noiseless map using moment mapping and clipping.

To generate a realistic spectral-line data cube of a molecular cloud that is, for our purpose, noiseless, we will utilize our extensive CO survey of the nearby Taurus-Auriga cloud complex (hereafter, the "Taurus cloud"; survey #21 in Table 1 of Dame et al. 2001); see Figure 1. Taurus is not a particularly large or massive molecular cloud, but it is very close and covers a large solid angle. Consequently more time was spent to beamwidth map this cloud—over 900 hours—than was spent on any other single cloud or complex in the CfA CO survey archive. By degrading the angular resolution of this survey by a factor of 8 and then re-sampling on an 8 times coarser grid, we can precisely simulate a survey of a cloud like Taurus that is 8 times further away and for which we integrated 64 times as long per point, roughly one hour. Doing this reduces the rms noise per point from 0.25 K to 0.03 K. Since this noise is negligible compared with the rms Gaussian noise of 0.5 K that we will subsequently add for our experiment, we will refer to this smoothed cube as "noiseless".

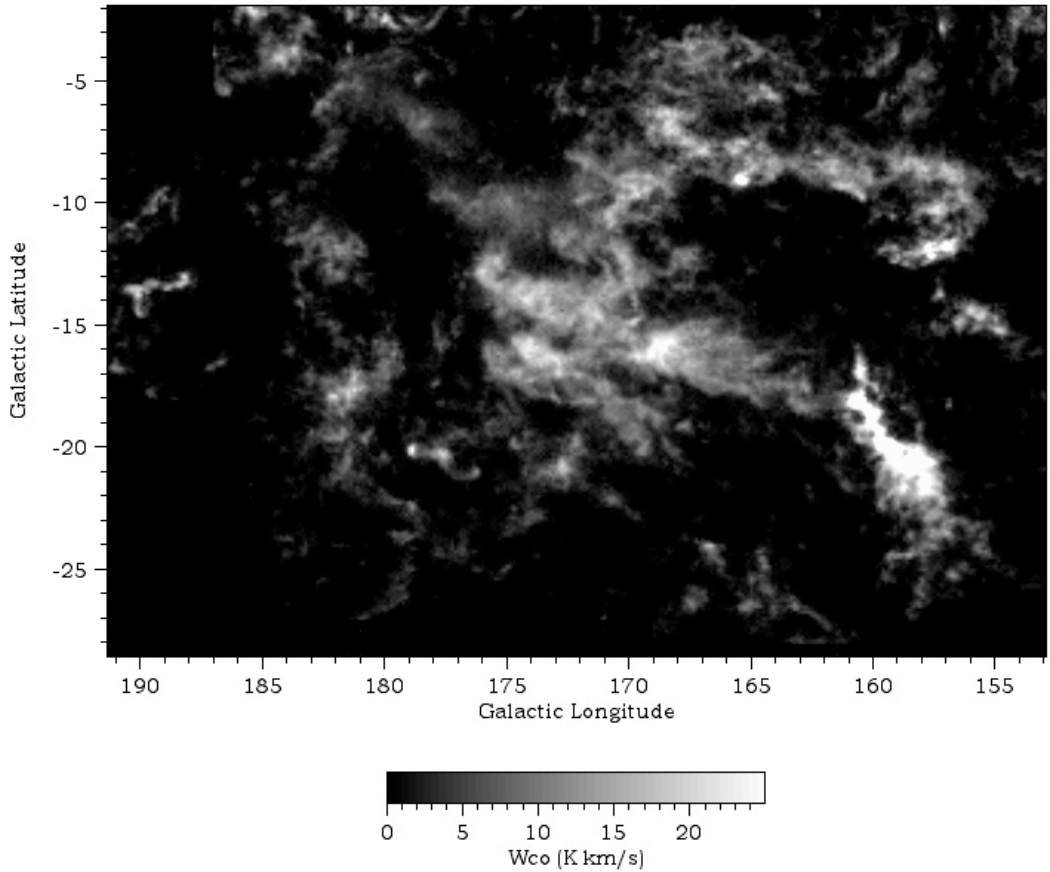

**Figure 1**: The Taurus-Auriga CO survey integrated over velocity. This is survey #21 from Table 1 of Dame et al. (2001).

We moved the noiseless Taurus survey into an empty data cube with larger velocity and spatial ranges: ±60 km/s in velocity, 50° in longitude by 42° in latitude. The extended ranges provide emission-free regions for displaying pure noise in the spectra and integrated maps. The extended velocity coverage is large for a survey of a local cloud, but comparable to the coverage of many of our Galactic plane surveys.

Integrating this noiseless survey over velocity yields the Wco map shown in Figure 2a. This is the map we will attempt to reproduce from a very noisy version of the same survey. A sample grid of the "noiseless" spectra is shown in Figure 2b. Adding Gaussian noise with an rms of 0.5 K to the cube and again integrating over velocity yields the Wco map in Figure 3a, with contour levels the same as in Figure 2a. The same array of sample spectra, now with the noise added, is shown in Figure 3b.

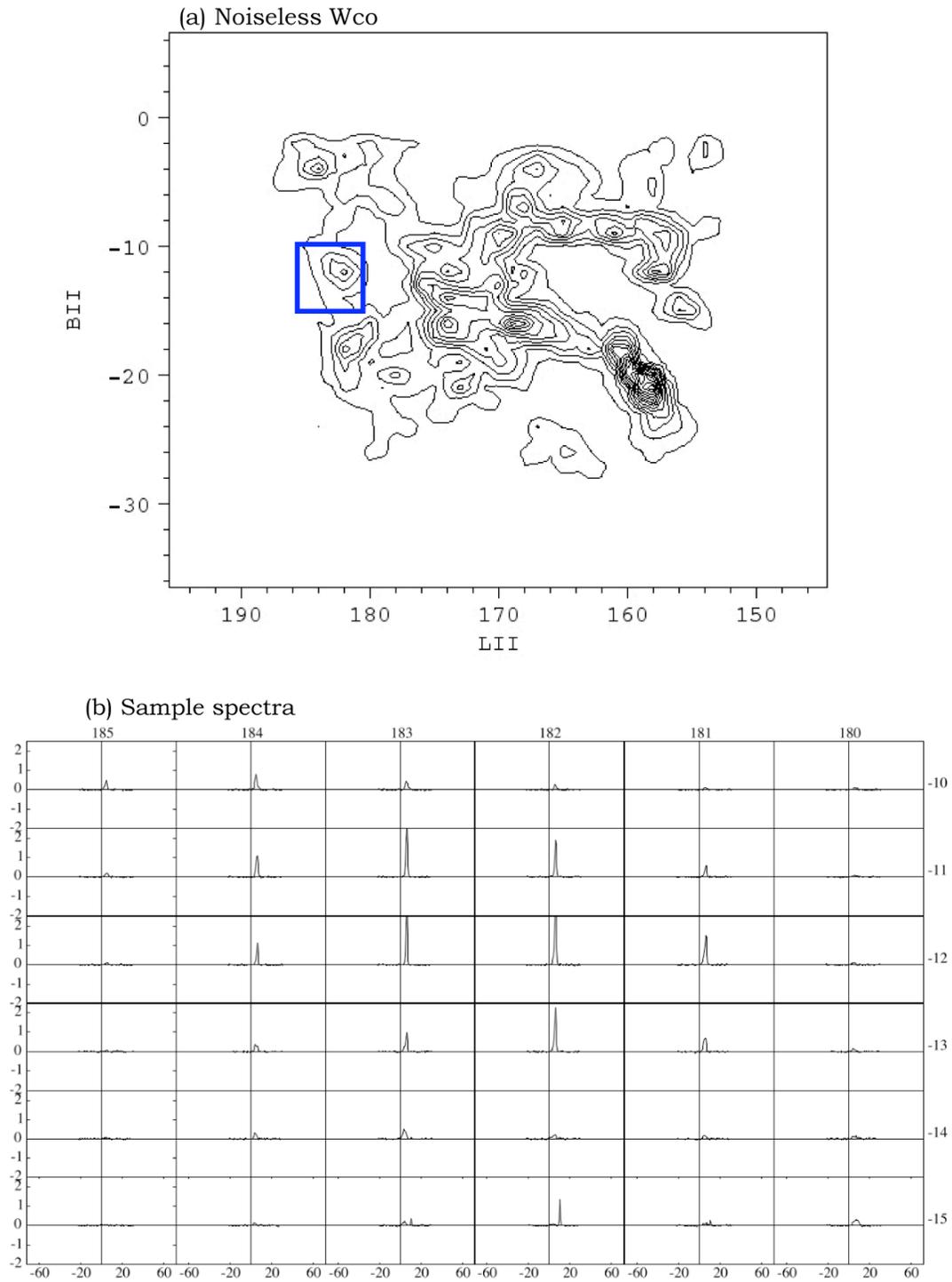

**Figure 2**: (a) Velocity-integrated CO map of the Taurus cloud after smoothing to a resolution of 1° and resampling on a 1° grid. Spectra within the region outlined in blue are shown in (b).

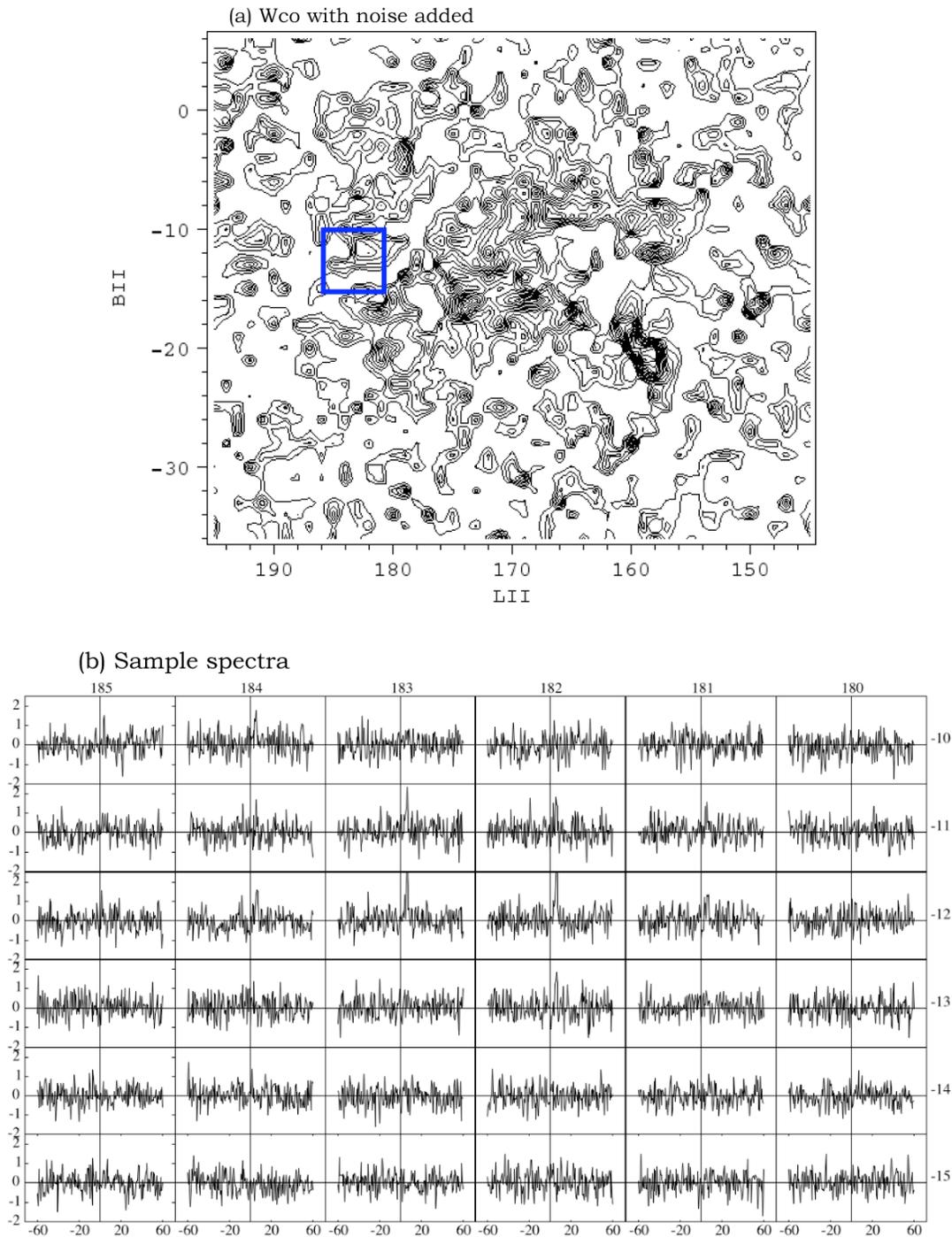

**Figure 3**: (a) Velocity-integrated CO map of the Taurus cloud after adding 0.5 K rms of Gaussian noise. Spectra within the region outlined by the blue box are shown in (b).

Next we generate clipped and moment masked versions of the noisy cube at various clipping and masking levels. Keep in mind that the sigma used for clipping is the raw rms of 0.5 K, whereas in moment masking sigma refers to the rms in the cube after the smoothing described in step ii above. The same sample spectra after this smoothing are shown in Figure 4a, and after moment masking at 5-sigma in Figure 4b.

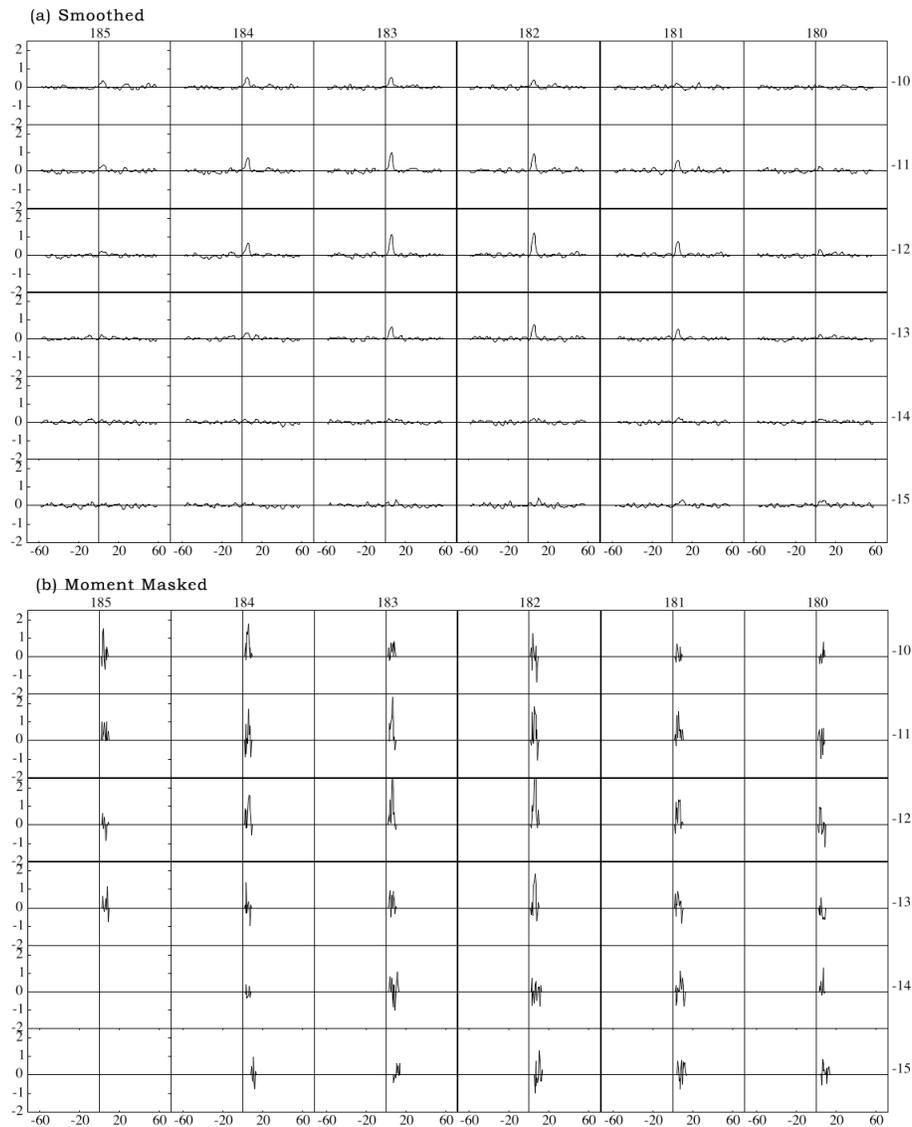

**Figure 4**: The same sample spectra as in Figs 2 & 3, but now smoothed as described in step ii of the moment masking procedure (a) and moment masked at 5-sigma (b).

The Wco maps obtained from moment masked and clipped cubes at various sigma levels are shown in the Figures 5 and 6 respectively. It's clear from a comparison of these maps with the noiseless map in Figure 2a that moment masking at 5-sigma (Fig. 5c) best reproduces the noiseless map, far better than clipping at any sigma level (Fig. 6). Moment masking below 5 sigma allows pure-noise features to enter the Wco maps while picking up little obvious additional emission. For the purpose of reducing noise there is little advantage to moment masking above 5-sigma, and doing so suppresses some of the weaker emission features (e.g., Fig 5d). For reference, the noiseless Wco map and the one moment masked at 5-sigma are reproduced side-by-side in Figure 7.

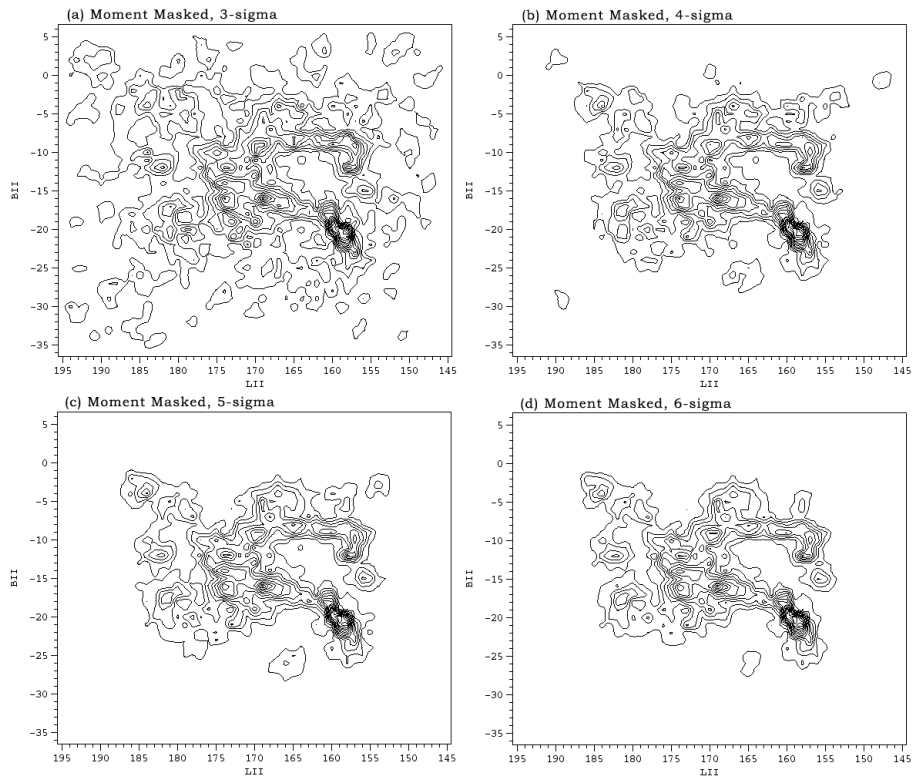

**Figure 5**: Moment masked CO maps with the masking level varying from 3-sigma in (a) to 6-sigma in (d).

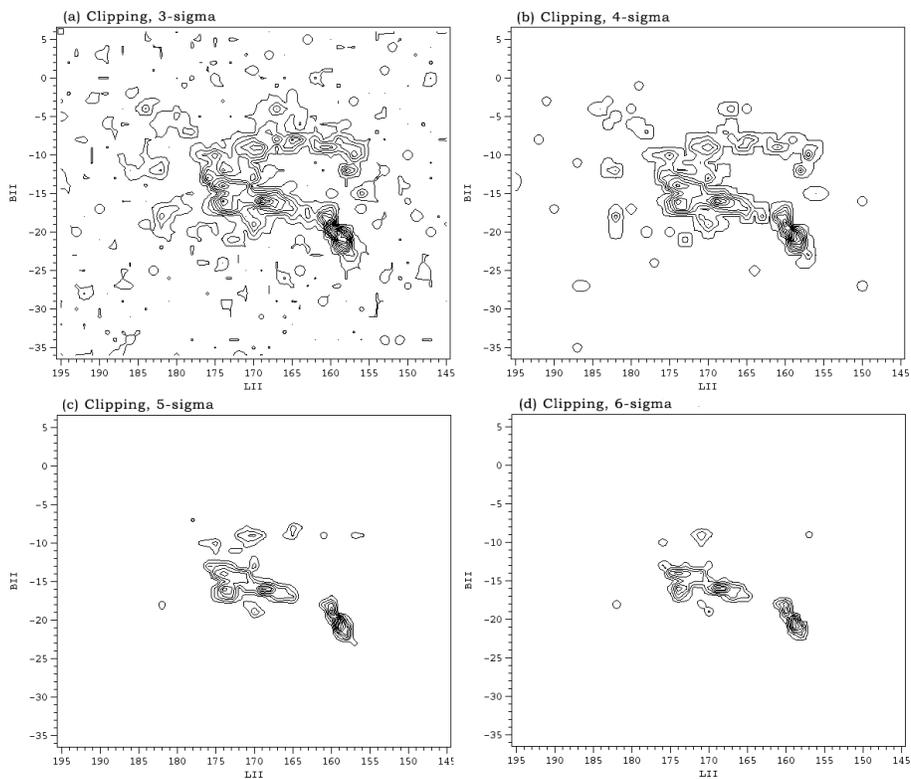

**Figure 6**: Clipped CO maps with the clipping level varying from 3-sigma in (a) to 6-sigma in (d).

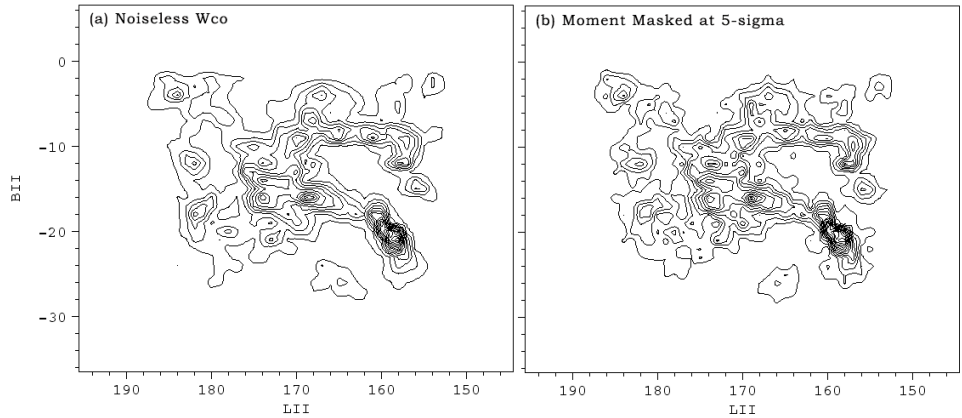

**Figure 7**: A comparison of the noiseless Wco map with that obtained by moment masking at 5-sigma.

Another criterion for determining the optimum noise suppression technique is to compare the total molecular cloud mass determined from each of the noise-suppressed cubes with that obtained from the noiseless cube. On the usual assumption that $H_2$ column density is proportional to CO velocity-integrated intensity, the total cloud mass is proportional to the sum over all channels in the cube; we will take that sum as the mass for the present purpose. In Figure 8 we plot the cloud mass as determined from cubes clipped and moment-masked over a wide range of sigma levels, and also the true mass determined from the noiseless cube.

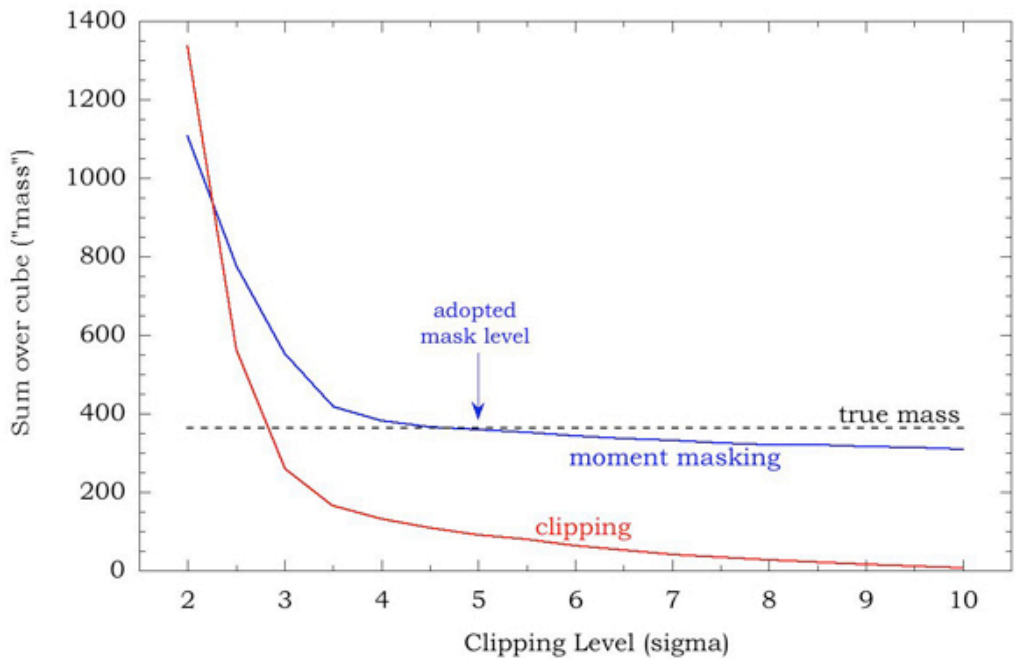

**Figure 8**: The total sum over clipped (red) and moment masked (blue) cubes for clipping and masking levels ranging from 2 to 10 sigma. The value of the sum over the noiseless cube is marked by a dashed line. The sum over the cube is proportional to the $H_2$ mass.

By this criterion also, we find that moment masking at 5-sigma is most effective at reproducing the noiseless result. The clipping technique at ~3-sigma also yields the correct mass but judging from Fig. 6a this is clearly owing to a positive noise bias. Both techniques suffer from this bias at low sigma, however clipping suppresses the bias only when significant emission is suppressed as well, whereas moment masking can suppress the bias while still maintaining most of the true emission. The moment-masking technique yields 88% of the true mass even when masking at 10 sigma, since most of the emission comes from large, bright features which have extremely high significance in the smoothed cube.

6. Summary

We have shown that for a typical molecular cloud such as that in Taurus, with bright central condensations, filaments, and weaker features around the periphery, the moment masking technique as refined here is highly effective in removing noise from the spectral line data cube while maintaining most of the emission at the full angular and spectral resolution of the original survey.

In applying the technique we smooth in velocity by only a factor of 2-4 in order to maintain resolution on the narrower spectral lines, and we smooth spatially by only a factor of 2. The spatial smoothing will diminish our sensitivity to single isolated emission peaks, but in a survey with millions of spectral channels there is, in any case, no way to confirm the reality of a single 3-4 sigma peak short of additional observations or reference to other surveys. Masking at 5-sigma in the smoothed cube ensures that no pure-noise features will pass into the moment maps, while our refinement of unmasking all pixels within the smoothing kernel of each 5-sigma peak in the smoothed map ensures that most of the emission in the raw cube is maintained.

In closing we emphasize that while we believe that the moment masking parameters advocated here are near optimum for analyzing the CfA CO surveys, they may not be for other surveys with very different S/N, resolutions, and emission structure.